\documentclass[aps,prl,twocolumn,superscriptaddress]{revtex4}
\usepackage{epsfig}
\usepackage{color}
\usepackage{hyperref}
\usepackage{graphicx}

\usepackage{amsmath}
\usepackage{comment}

\newcommand{\EqLabel}[1]{\label{#1}}

\begin{document}
 
\title{The critical nature of the Ni spin state in doped NdNiO$_2$} 

\author{Mi Jiang} 
\affiliation{Department of Physics and Astronomy, University of
British Columbia, Vancouver B.C. V6T 1Z1, Canada} 
\affiliation{Stewart Blusson Quantum Matter Institute, University of
British Columbia, Vancouver B.C. V6T 1Z4, Canada}

\author{Mona Berciu}
\affiliation{Department of Physics and Astronomy, University of
British Columbia, Vancouver B.C. V6T 1Z1, Canada}
\affiliation{Stewart Blusson Quantum Matter Institute, University of
  British Columbia, Vancouver B.C. V6T 1Z4, Canada}

\author{George A. Sawatzky}
\affiliation{Department of Physics and Astronomy, University of
British Columbia, Vancouver B.C. V6T 1Z1, Canada}
\affiliation{Stewart Blusson Quantum Matter Institute, University of
British Columbia, Vancouver B.C. V6T 1Z4, Canada}

\begin{abstract} % less than 600 characters including spaces
Superconductivity with $T_c \approx 15K$ was recently found in doped
NdNiO$_2$. The Ni$^{1+}$O$_2$ layers are expected to be Mott insulators so
hole doping should produce Ni$^{2+}$ with $S=1$, incompatible with
robust superconductivity. We show that the NiO$_2$ layers fall
inside a ``critical'' region where the large $pd$ hybridization favors a
singlet $^1\!A_1$ hole-doped state like in  CuO$_2$. However, we find
that the superexchange is about one order smaller than in
cuprates, thus a  magnon ``glue'' is very unlikely and another mechanism needs to be found.
\end{abstract}

\maketitle

{\em Introduction:} Understanding the mechanism responsible for the
appearance of high-temperature superconductivity (SC) in
cuprates~\cite{BedMu} remains one of the top challenges in condensed
matter physics. Despite over 30 years of intense effort that produced
numerous proposals, there is no consensus on what is the ``pairing
glue'' ~\cite{Anderson2007}.

\begin{figure}[b]
 \includegraphics[width=0.9\columnwidth]{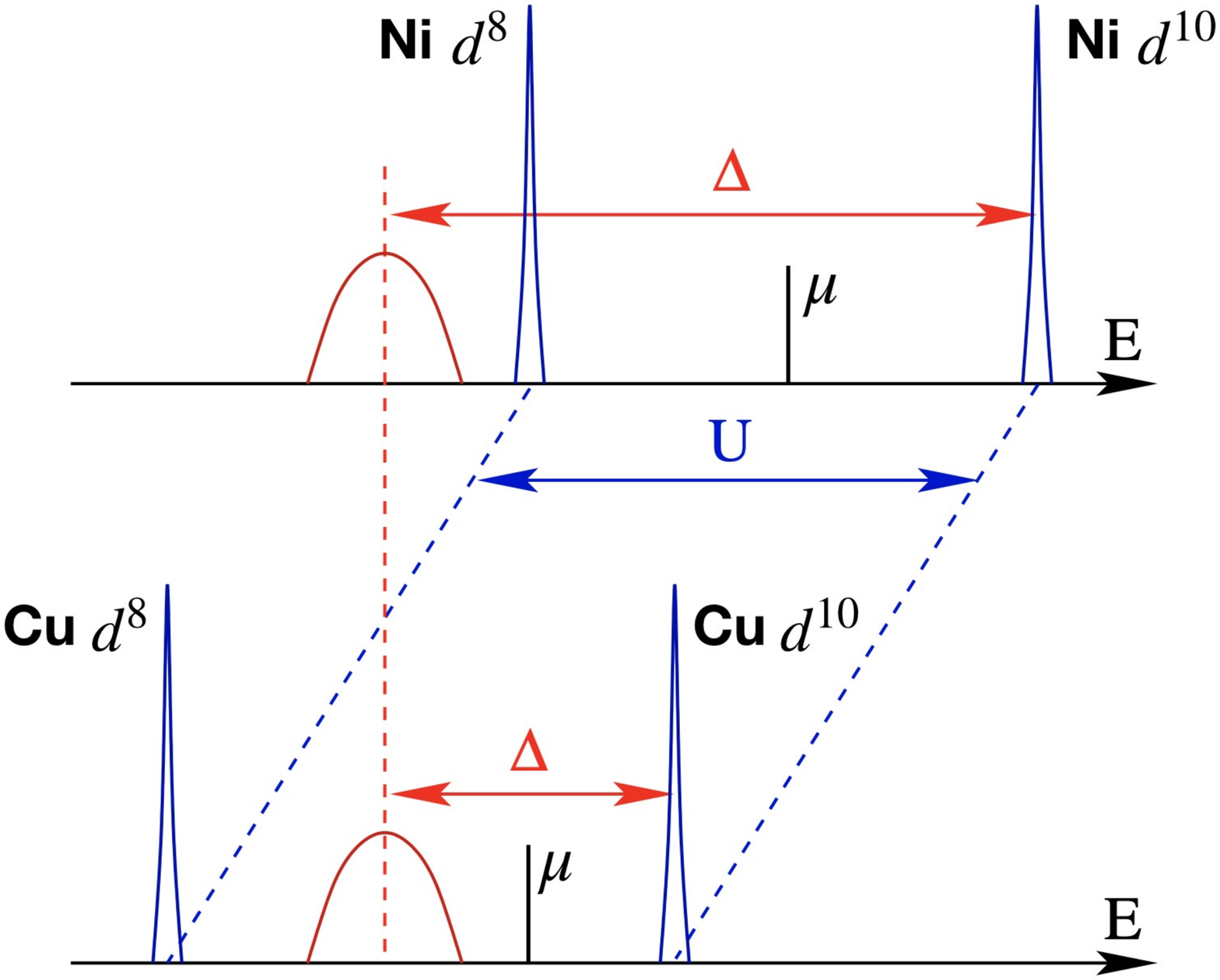}
 \caption{(color online) Sketch of a Mott insulator (top) {\em vs.} a
   charge transfer insulator (bottom). The narrow (blue) bands are the
   Hubbard $3d$ bands while the broader (red) band is the O$2p$ band {\em before}
   the $pd$ hybridization has been switched on. The
   sketch assumes a similar $U$ but a significantly larger $\Delta$,
   like for NdNiO$_2$ as compared to LaCuO$_4$. Below (above)
   the chemical potential $\mu$ are the electron removal (addition)
   states, if the starting configuration is Ni/Cu $3d^9$ plus a full
   O $2p$ band.}
\label{fig1}
\end{figure}

A possible route of breaking this deadlock is to find similar but
non-Cu based families of SC, which might help solve the mystery. One
of the routes being pursued is to replace Cu$^{2+}$ with Ni$^{1+}$ in
compounds such as LaNiO$_2$ and NdNiO$_2$. Both Cu$^{2+}$ and
Ni$^{1+}$ are in the $3d^9$, $S={1\over 2}$ configuration in the
parent compound, so the infinite NiO$_2$ planes appear to be direct
counterparts of the CuO$_2$ planes. After several failed
attempts~\cite{LaNiO2a,LaNiO2b,LaNiO2c}, SC with $T_c$ of up to 15K
was recently found in doped Nd$_{0.8}$Sr$_{0.2}$NiO$_2$ single crystal
thin films~\cite{2019Nature}. This is a very exciting development,
suggesting the existence of a Ni-based family of high-$T_c$ SC that
may be more like the cuprates than the Fe-based
family~\cite{Honoso2006} turned out to be.

To gauge how similar this new SC is to cuprates, we compare the
behaviour of a NiO$_2$ layer upon doping to that of a CuO$_2$ layer.
Throughout this work, we assume that, like in CuO$_2$ layers, only the
O $2p$ and the Ni $3d$ states determine the low-energy physics of the
NiO$_2$ layer, and that stochiometric NiO$_2$ layer is a large gap
insulator. We note that the nature of the parent compound NdNiO$_2$ is
not yet clear. The thin film in Ref. \cite{2019Nature} is metallic,
but recent reports found both thin films and bulk crystals that are
insulators \cite{B1,B2}. As further discussed below, ab-initio studies
are also divided; those that find a metallic band, locate it in the Nd
layers. If this Nd band exists, charge neutrality demands that electrons present in the Nd layers in the stoichiometric compound, be compensated by holes in the NiO$_2$ layers, further emphasizing the need to understand the latter subsystem.

For all of these reasons, here we study a NiO$_2$ layer. A major
difference is immediately apparent: NiO$_2$ should be a Mott
insulator~\cite{Hepting,Jinlong,Arita1,Cano1} while the cuprates are
charge-transfer insulators, according to the Zaanen-Sawatzky-Allen
(ZSA) scheme~\cite{ZSA1985}, see sketch in Fig. \ref{fig1}. This is
due to a charge transfer energy $\Delta \approx 9$ eV in NiO$_2$ {\em vs.} $\Delta \approx3$
eV in CuO$_2$, because the smaller nuclear charge of Ni causes a 5-6eV
upward shift of the $d^{10}$ state in NiO$_2$ (the other energy
scales, in particular $U_{dd}\approx 6-7$eV, are similar in both types of
layers, see below)~\cite{Sawatzky1990}.

Holes doped in a Mott insulating NiO$_2$ layer would reside on the Ni,
not in the O$2p$ band. Because Ni$^{2+} (3d^8)$ is $S=1$ in all other
known Ni$^{2+}$ oxides, this would make the appearance of rather
high-T$_c$ superconductivity very puzzling, and definitely unlike that
in cuprates.

In this Letter, we use an impurity calculation including the full
Ni:$3d^8$ multiplet structure to argue that in fact, NiO$_2$ lies in
the critical crossover region of the ZSA diagram, where the lowest
$S=0$ and $S=1$ eigenstates cross. It is possible, therefore, that
holes doped in NiO$_2$ layers have a strong O $2p$ component with the
same $^1\!A_1$ symmetry like the Zhang-Rice singlet (ZRS) of cuprates,
{\em i.e.} two holes in orbitals with $x^2-y^2$
symmetry~\cite{ZhangRice}. This may explain how SC could emerge upon
doping of this Mott insulator, as it would make hole doping in NiO$_2$
rather similar to cuprates. It would also suggest that small changes
in the parameters, resulting from variations in lattice parameters
through chemical composition or applied pressure, could also stabilize
the more common $^3\!B_1$ lowest energy hole state, {\em i.e.} one
holes in $x^2-y^2$ and the other in $3z^2-r^2$ symmetry. This is a
testable prediction, which may be linked to the lack of SC in the
closely related LaNiO$_2$ compound~\cite{2019Nature}. It is also
consistent with the reported differences in the properties of thin
films and bulk crystals grown in different labs, and their high
sensitivity to applied pressure.

While a $^1\!A_1$, ZRS-like type of hole-doped state would make
NiO$_2$ similar to CuO$_2$, it does not automatically confirm a
similar SC like in cuprates. This is because another consequence of
the larger $\Delta$ is that the superexchange~\cite{superexchange}:
$$ J_{dd} = \frac{4t_{pd}^4}{\Delta^2 U_{dd}}
+\frac{8t_{pd}^4}{\Delta^2(U_{pp}+2\Delta)}
  $$ must be around ten times smaller in NiO$_2$ than in
cuprates~\cite{Jinlong}. While this explains the lack of
antiferromagnetic (AFM) order in the parent
compound~\cite{LaNiO2b,LaNiO2c}, it also makes it very unlikely that
SC in the new material is mediated through magnon-exchange. Insofar as
that is a leading scenario for cuprate SC, this means either that the
new SC has a different mechanism than cuprates, or that SC in cuprates
is not (primarily) magnon-mediated. We further discuss these
implications below.

{\em Model:} We study a hole doped into a system consisting of a
Ni$^{1+}$ ($3d^9$) impurity properly embedded in an infinite square
lattice of O$2p^6$ ions. The Hamiltonian is:
%%%%%%%%%%%%%%%%%%%%%%%%%%%%%% EQUATION %%%%%%%%%%%%%%%%%%%%%%%%%%%%%%
\begin{equation}
\EqLabel{Ham} {\cal H} =
\hat{U}_{dd}+\hat{T}_{pd}+\hat{T}_{pp}+\hat{\Delta} + \hat{U}_{pp}.
\end{equation}
%%%%%%%%%%%%%%%%%%%%%%%%%%%%%%%%%%%%%%%%%%%%%%%%%%%%%%%%%%%%%%%%%%%%%%
Here, $\hat{U}_{dd}$ includes all Coulomb and exchange integrals of
the $3d^8$ multiplet, which determine the spin and symmetry of the
lowest-energy hole addition state. Matrix elements between the $3d$
orbitals $b_1(d_{x^2-y^{2}})$, $a_1(d_{3z^2-r^{2}}), b_2(d_{xy}),
e_x(d_{xz})$ and $ e_y(d_{yz})$ are in terms of the Racah parameters
$A,B,C$~\cite{Racah}. The $3d$ orbitals are assumed to be degenerate, {\em i.e.} we
omit point-charge crystal splittings. (This is a good approximation
because it is the hybridization with the O orbitals, included in our
model, that accounts for most of the difference between the effective
on-site energies of the $3d$ levels). $ \hat{T}_{pd}$ and
$\hat{T}_{pp}$ describe hopping of holes between the Ni$3d$ orbitals
and adjacent O$2p$ ligand orbitals, and between nearest neighbour O
ligand orbitals, respectively. We note that we have checked explicitly
that including the second pair of in-plane, $\pi$-type O$2p$ orbitals
has essentially no effect on the results reported below. Finally,
$\hat{\Delta}$ measures the difference in the on-site energies of $3d$ and
$2p$ orbitals, as measured from the full O 2p band to the Ni $d^{10}$
state, while $\hat{U}_{pp}$ is an on-site Hubbard repulsion when two
holes are in the same O ligand orbital.

\begin{figure}[t]
\includegraphics[width=\columnwidth]{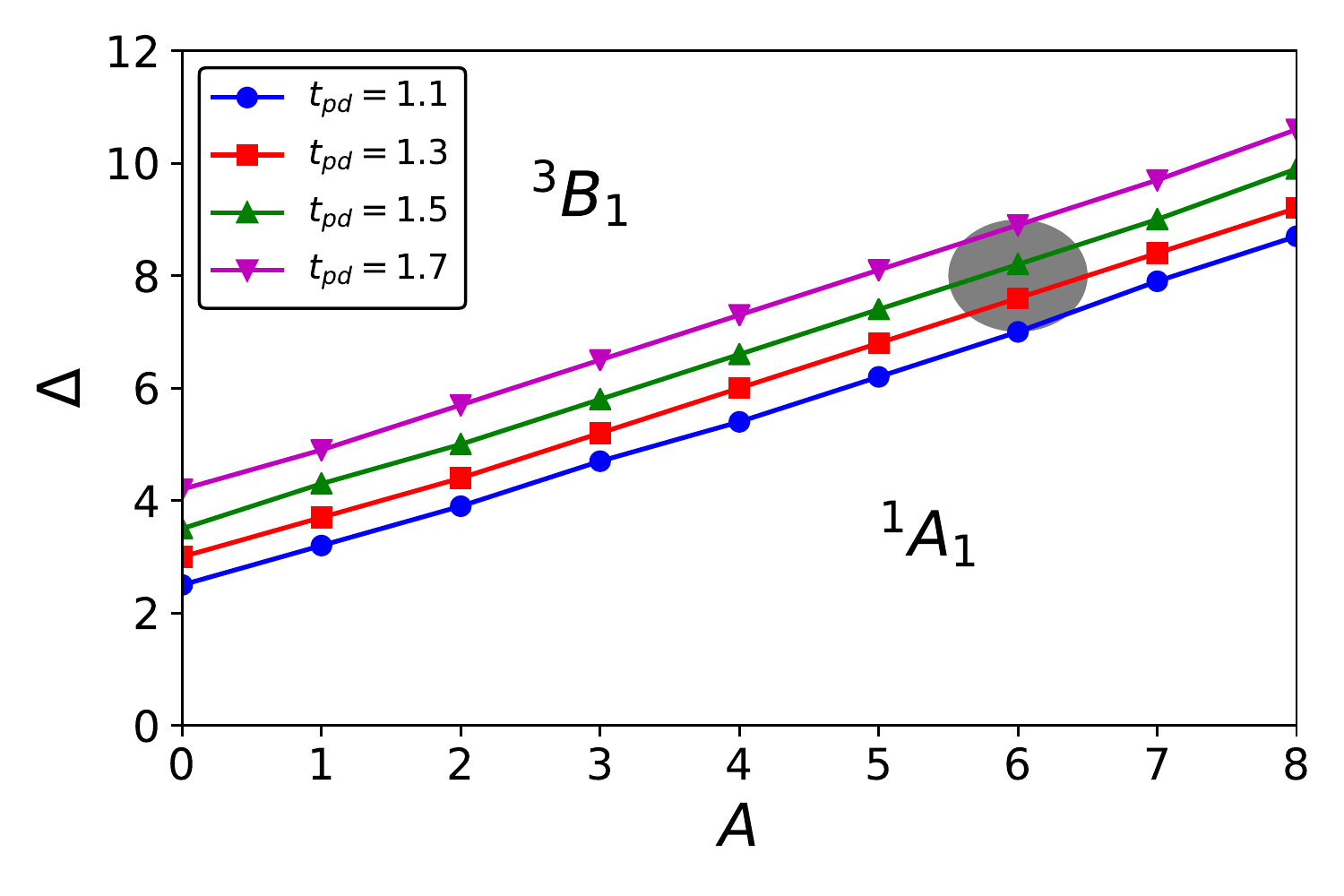}
\caption{(color online) Phase diagram for the stability of a $^3\!B_1$ triplet ground-state, expected for a doped Mott insulator, {\em vs.}  a $^1\!A_1$ singlet ground-state like in hole-doped cuprates, for different values of $\Delta$ and $A$, measured in eV. Other parameters are $t_{pp}=0.55eV, B=0.15eV, C=0.58eV, U_{pp}=0$. The four lines correspond to $t_{pd}=1.1, 1.3, 1.5$ and $1.7$eV. We expect NdNiO$_2$ to lie in the shaded region.}
\label{fig2}
\end{figure}

\begin{figure*}
\psfig{figure=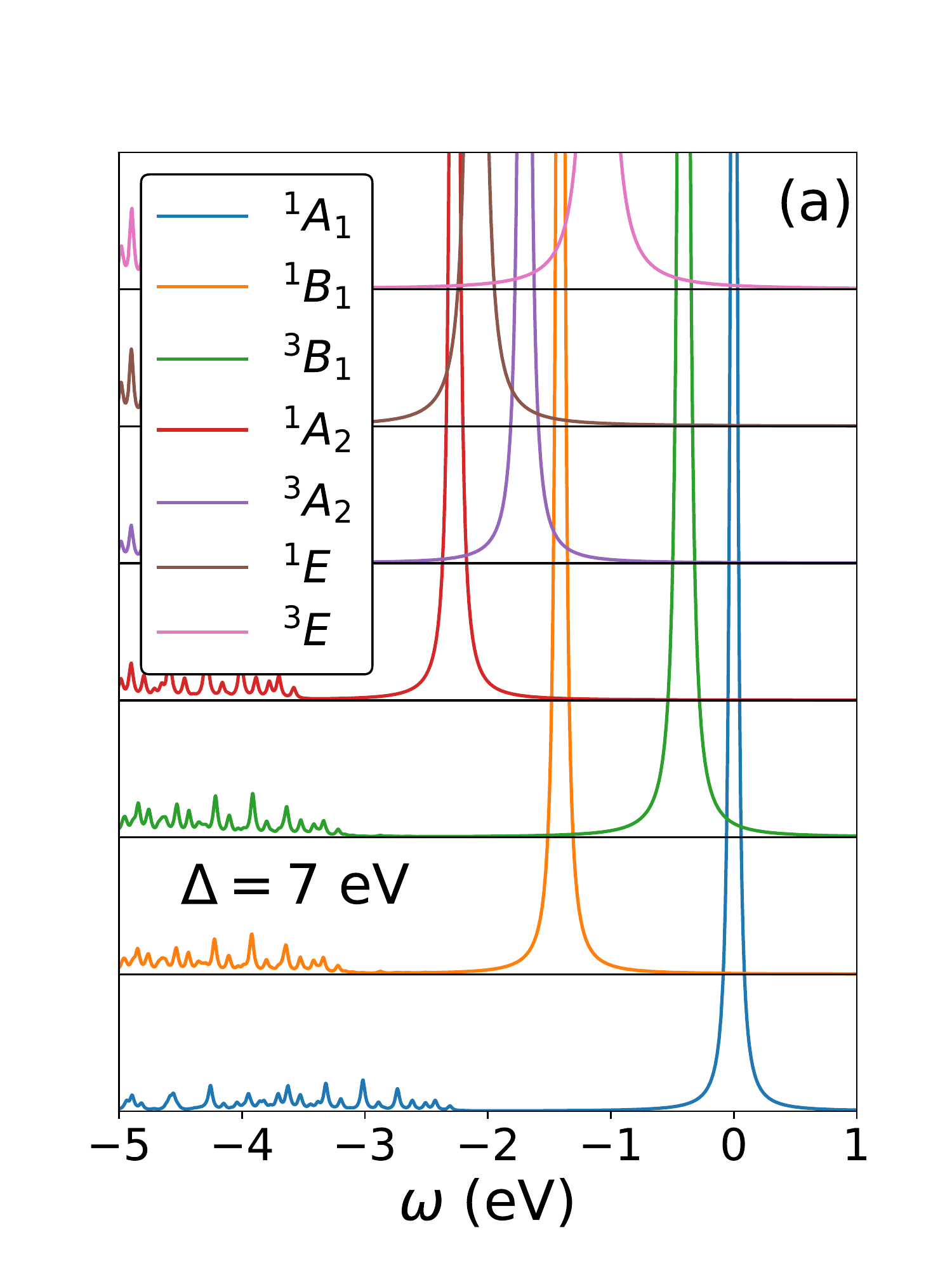,height=7.8cm,width=.32\textwidth, clip=true, trim = 0.7cm 0.0cm 1.3cm 0.0cm}
\psfig{figure=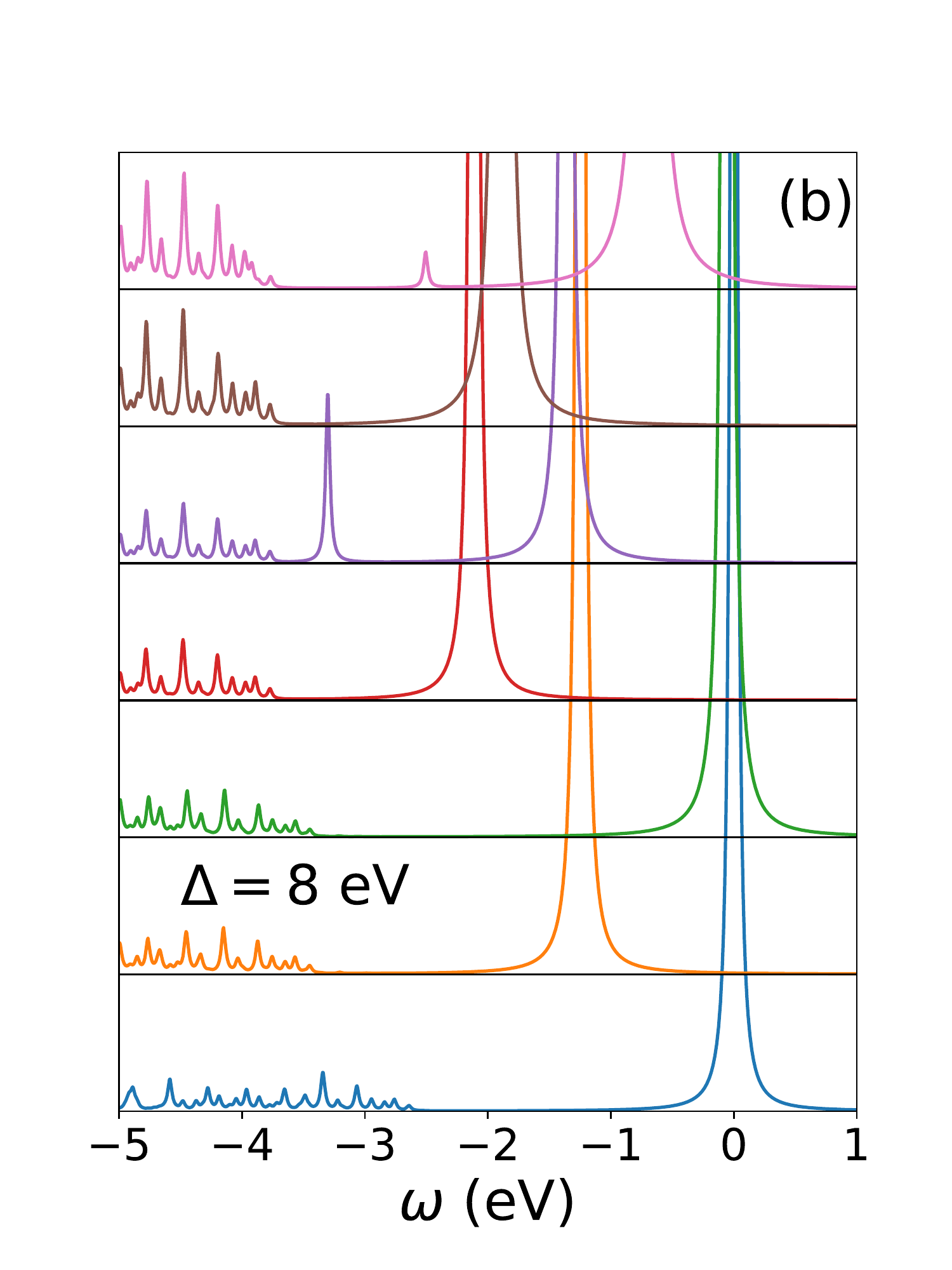,height=7.8cm,width=.32\textwidth, clip=true, trim = 1.0cm 0.0cm 1.0cm 0.0cm} 
\psfig{figure=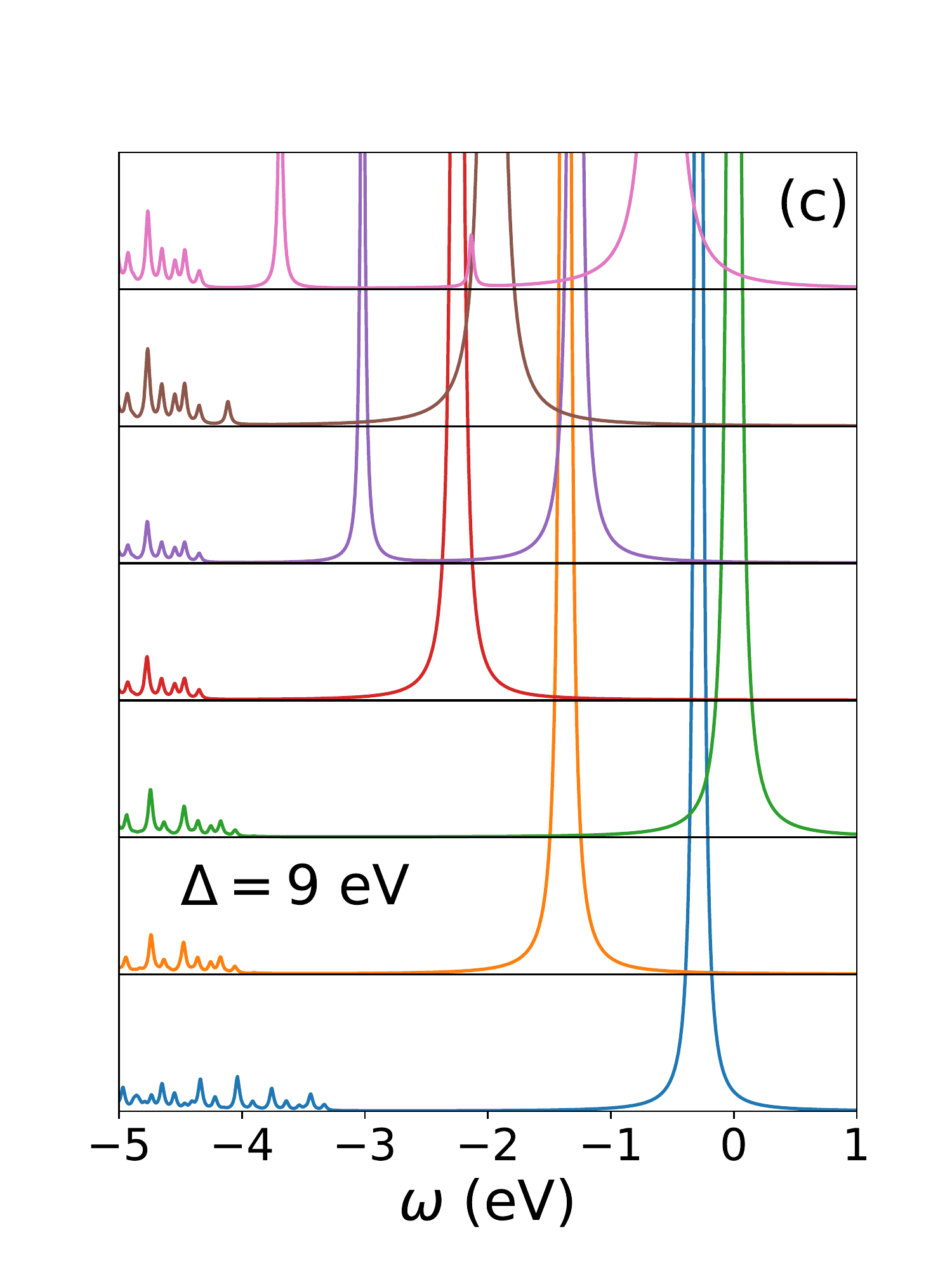,height=7.8cm,width=.32\textwidth, clip=true, trim = 1.3cm 0.0cm 0.7cm 0.0cm}  
\caption{(color online) Spectral weight for addition of a doped hole, when $t_{pd}=1.5$eV and $A=6.0$eV are kept constant, while $\Delta= 7, 8, 9$eV in the left, central and right panels, respectively. Spectra are shifted such that the ground-state is located at zero energy. With increasing $\Delta$, its character changes from $^1\!A_1$ (spectrum shown by the blue, lowest curve) to $^3\!B_1$ (spectrum shown by the green, 3rd lowest curve). }
\label{fig3}
\end{figure*}

We study the GS of this impurity system with either one hole (undoped)
or two holes (doped) using Exact Diagonalization. We recently used
this model to study a Cu$^{2+}$ impurity embedded in a square O
lattice~\cite{Mi2019}; we refer the reader there for further details, in particular why the proper modelling of the O sublattice is essential.\cite{noteRefD}
We use fits to ab-initio results~\cite{Kat,Held,Millis} to extract the hybridization 
  between the Ni impurity and neighbor O,  $t_{pd}\approx 1.3-1.5$eV, and  between adjacent O,  $t_{pp}\approx0.55$eV. These values are similar to those in the
  cuprates. For $t_{pd}$, this  is because of a partial
  cancellation of changes due to the somewhat
larger lattice constant of the nickelates, and to the larger orbital
radius of their $d$ orbital, due to the smaller nuclear charge.
For $t_{pp}$, this is because the lattice constant increase is not that
significant. We keep the same values for the Racah parameters $B, C$,
because they are set by atomic physics and not much influenced by
screening effects, as verified experimentally in other systems with
Ni$^{2+}$ ions~\cite{Sawatzky1990}. $U_{pp}$ plays little role for the
results discussed below, so we set it $U_{pp}=0$ (very similar results
are obtained for $U_{pp}\approx 3$eV, however). As already mentioned, we
expect the on-site Coulomb repulsion on the Ni$^{1+}$ to be comparable
to that on the Cu$^{2+}$, $U_{dd}=A+8B+3C\approx 6-7$eV, while the charge
transfer $\Delta\approx 7-9$eV as opposed to $\Delta\approx 3$eV in
cuprates~\cite{Christensen}.

{\em Results:} Figure \ref{fig2} shows the one-hole phase diagram as a
function of $A$ and $\Delta$. It consists of two regions with
ground-states (GS) of $^1\!A_1$ and $^3\!B_1$ symmetry~\cite{Mi2019,Ballhausen,Eskes}, respectively. In
the $^3\!B_1$ region the doped hole primarily sits on the Ni and locks
into a triplet with the other hole, by Hund's exchange. More
specifically (also see below), one hole occupies a wavefunction with
dominant $3d_{x^2-y^2}$ character plus a small contribution from the
$x^2-y^2$ linear combination of adjacent O$2p$ orbitals, while the
second hole occupies the $3z^2-r^2$ counterpart. In
contrast, in the $^1\!A_1$ region the doped hole occupies primarily
the $x^2-y^2$ ``molecular''-like O$2p$ orbital with a small admixture
of the $3d_{x^2-y^2}$ orbital, and is locked in a singlet with the
other hole which has primarily $3d_{x^2-y^2}$ character.  

The three lines show how the boundary shifts with
$t_{pd}$, and the shaded ellipse is the area we believe to
be relevant for the NiO$_2$ layer.  Clearly, it falls in the
borderline regime where the ground-state changes its nature.

To better understand what happens, and how it is possible for the doped states of a a Mott
insulator to instead look more like those of a charge-transfer insulator, we
plot in Fig. \ref{fig3} spectral densities for the addition of the
doped hole, resolved by point symmetry. The three panels correspond to
$\Delta=7,8,9$eV while $t_{pd}=1.5eV$ and $A=6.0$eV. They show that
with increasing $\Delta$, the ground-state transitions from having
$^1\!A_1$ symmetry to having the $^3\!B_1$ symmetry expected in the
large $\Delta$ limit. (The same transition is seen varying
$t_{pd}$ for a fixed $\Delta$.)

With increasing $\Delta$, we see an
increasing separation between the discrete peaks showing the
low-energy states with the doped hole bound to the impurity Ni with
various symmetries (these include the ground-state, always shifted to
occur at zero energy), and the continuum which describes excited states
with the doped hole moving freely in the O band. (The band is a
sequence of closely spaced peaks because for computational
convenience, we limit the size of the $O$ lattice to be $20\times20$.
In the thermodynamic limit, this consequence of confinement to a
``finite-box'' disappears and the continuum becomes smooth, but its
band-edges do not move).

Physically, what happens is that when the separation between the
$3d^8$ peak (which depends on the symmetry and the spin of the state)
and the O band sketched in Fig. \ref{fig1} is smaller or
comparable with $t_{pd}$, their hybridization is responsible for the
appearance of these low-energy bound states in all the channels; in
particular, the one with $^1\!A_1$ symmetry as the ground-state. As
$\Delta$ increases and the distance between the two features becomes
larger than $t_{pd}$, the ground-state switches to the expected
$^3\!B_1$ character for a Mott insulator, which is the conventional Hund's rule ground state for Ni $d^8$ .

\begin{figure}
\includegraphics[width=\columnwidth]{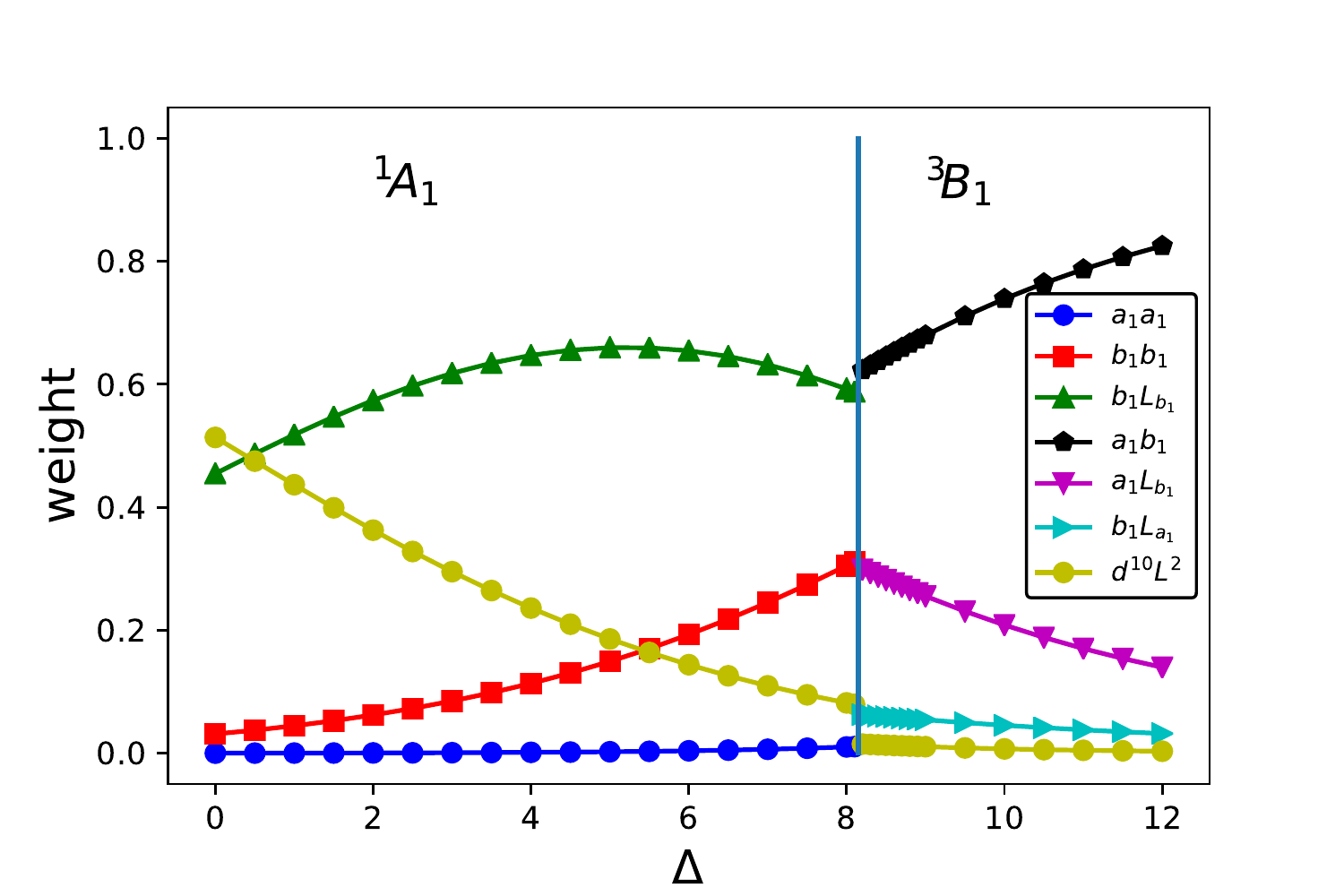}
\caption{(color online) Variation of the ground state weights of the
dominant components versus $\Delta$, for fixed $t_{pd}=1.5$ eV,$A=6.0$ eV. Note that the dominant component changes dramatically as the ground state switches from $^1\!A_1$ to $^3\!B_1$. The vertical line denotes the critical value $\Delta=8.1$ eV separating the two phases.}
\label{fig4}
\end{figure}

 In Figure
~\ref{fig4} we show the evolution of the ground state weights of the various
components with increasing $\Delta$, for  $t_{pd}=1.5$eV  and $A=6$eV. 
These components of the two-hole wavefunction are products of
the single particle states of various symmetries. Included in this
analysis are all the $d^8, d^9L$ and $d^{10}L^2$ states. The weights
of the states involving the hole continua are summed over energy. 

We see that the only $d^8$ basis states with reasonable weights are
the $a_1a_1,b_1b_1$, and $a_1b_1$ states (we use the standard notation $a_1 \equiv d_{3z^2-r^2}, b_1\equiv d_{x^2-y^2} $\cite{Mi2019,Eskes}). These are also the only
symmetries that are important for the $d^9L$ and $d^{10}L^2$ continuum
states. We see that for $\Delta<A$, {\em i.e.} in the charge transfer
gap region, the amount of total $d^8$ character is  small, 
increasing as we approach
the transition to the $^3\!B_1$ ground state. Simultaneously, the
amount of $d^{10}L^2$ is gradually decreasing because the separation
between the $d^{10}L^2$ and the $d^9L$ continua is increasing as
$\Delta$ increases. The transition occurs at $\Delta =8.1$eV, which is
considerably higher than where the transition from the Mott to the
charge transfer insulator occurs in the ZSA diagram. This is because
the $L_{b_1}$ symmetry states in the O 2p band are located at the top
of the O band and not at its center, so their effective $\Delta$ is
about 1.5 eV lower. Note the abrupt change in the weights of the $d^8$
components, especially the $a_1b_1$ component, as we enter the region
where the ground-state is $^3\!B_1$. Here, $d^8 (a_1b_1)$ is the
dominant component, unlike the ZRS-like $d^9(b_1)L_{b_1}$ dominating
the $^1\!A_1$ symmetry.

These results suggest that the minimal model able to describe both phases is the counterpart of Eq. (\ref{Ham}) that includes only the $e_g$ orbitals (not all five Ni $3d$ orbitals). Simpler minimal models are available to describe each phase. On the $^1\!A_1$ side, one can project down to a single band Hubbard-like or $t$-$J$-like model~\cite{ZhangRice}, although it is controversial whether the O orbitals should be integrated out~\cite{Bayo, Hadi1, Hadi2}. On the $^3\!B_1$ side, a so-called type-II $t$-$J$ model has been recently proposed to describe the motion of a hole with spin $S=1$ in a background of spin-${1\over 2}$ \cite{Ash}; older work along similar lines was reported in Ref. \cite{Oles}.

{\em Summary and discussion:} To conclude, we performed an impurity
calculation to demonstrate that despite nominally being in the Mott
region of the ZSA scheme, in fact a NiO$_2$ layer falls inside a
critical region of the parameter space where the strong $pd$
hybridization may favor a $S = 0$ hole-doped state with $^1\!A_1$
symmetry, similar to the cuprates. However, the triplet $^3\!B_1$
state is close in energy so that small changes in the parameters
resulting from changes in the lattice structure with chemical
substitution, with epitaxial strain or with pressure could stabilize the triplet state, making SC unlikely.

We also pointed out that these NiO$_2$ layers have large charge transfer energies $\Delta$, leading to a superexchange around an order of magnitude smaller than in cuprates. This fact severely challenges the scenario of spin fluctuations as the glue for superconductivity here, although that is currently the prevalent scenario in cuprates.

Before realistic proposals for the mechanism of
  superconductivity can be put forward, it is imperative to know
  whether metallic bands appear in the Nd layer. As mentioned, reports
  both of metallic and of insulating stoichiometric samples exist in
  the literature. Ab-initio studies are also divided. Some suggest
  that the parent compound is a metal with a Nd band crossing the
  Fermi energy \cite{Pickett,Botana,Kuroki,Weng,Han,Zhong,Arita,Chen,Lechermann,Jinlong,Millis},
  but this becomes gapped if correlations are included in a LDA+U
  approach~\cite{Kat}, and/or upon various chemical substitutions
  \cite{Kat,Held}. In any event, if metallicity due to bands located
  in the other layers is confirmed, it will certainly be necessary to
  understand their interplay with the NiO$_2$ layers. This would make
  superconductivity in the new compound definitely unlike that of
  cuprates, and would suggest the fascinating possibility of another
  mechanism that stabilizes high-T$_c$ superconductivity. While such
  work has already begun, see for {\em e.g.} Refs.~\cite{Hepting,Thomale,Mei,GuangMing,Ash,Hirsch,Werner,Congjun,
Cano,Mei1,Pickett1,Ding,Talantsev,Held1,Chunjing},  it is essential to first have a good undertanding of the individual
  behaviour of either kind of layer with doping. The work presented
  here demonstrates that the NiO$_2$ layers should not be treated like
  simple Mott insulators or charge transfer insulators.

\begin{acknowledgments}
 {\em Acknowledgements:}  This work was funded by the Stewart Blusson Quantum Matter Institute at University of British Columbia, and by the Natural Sciences and
  Engineering Research Council of Canada.
\end{acknowledgments}


\begin{thebibliography}{27}
\bibitem{noteRefD} Note that the relevant energy scale is the bandwidth $8 t_{pp} \approx 4.5$ eV. This is a considerable energy, which is why the proper modelling of the O band is important.
\end{thebibliography}
\end{document}